\documentclass[letterpaper]{mn2e}
\usepackage{psfig}
\usepackage[totalwidth=480pt, totalheight=680pt]{geometry}

\catcode`\@=11
\def\gsim{\ifmmode{\,\mathrel{\mathpalette\@versim>\,}}
    \else{$\,\mathrel{\mathpalette\@versim>}\,$}\fi}
\def\lsim{\ifmmode{\,\mathrel{\mathpalette\@versim<\,}}
    \else{$\,\mathrel{\mathpalette\@versim<}\,$}\fi}
\def\@versim#1#2{\lower 2.9truept \vbox{\baselineskip 0pt \lineskip
    0.5truept \ialign{$\m@th#1\hfil##\hfil$\crcr#2\crcr\sim\crcr}}}
\catcode`\@=12  
\def\kb{k_{\rm B}}

\def\me{m_e}
\def\tmin{T_{\rm in}}
\def\tmax{T_{\rm out}}

\def\tcool{t_{\rm cool}}
\def\tkappa{t_{\rm \kappa}}
\def\rmin{R_{\rm in}}
\def\rmax{R_{\rm out}}
\def\lcrit{l_{\rm crit}}

\def\taumin{\tau_{\rm in}}
\def\halpha{{\rm H}{\alpha}}

\def\Lam23{\Lambda_{23}}
\def\Lamtmax{\tilde{\Lambda}_{\tmax}}

\def\ximin{\xi_{\rm in}}

\def\ne{n_e}

\def\press{P}
\def\p6{P_6}
\def\qsat{q_{\rm sat}}
\def\qcl{q_{\rm cl}}
\def\ximin{\xi_{\rm in}}
\def\cmm3{\,{\rm cm}^{-3}}
\def\cm{\,{\rm cm}}
\def\kms{\,{\rm km}\,{\rm s}^{-1}}
\def\Myr{\,{\rm Myr}}
\def\sm1{\,{\rm s}^{-1}}
\def\kelvin{\,{\rm K}}
\def\kpc{\,{\rm kpc}}
\def\log{\,{\rm log}}
\def\ergs{\,{\rm erg}}
\def\kspitzer{\kappa_{\rm 0}}

\title[Cold filaments in galaxy clusters: effects of heat
conduction]{Cold filaments in galaxy clusters: effects of heat
conduction}

\author[C. Nipoti and J. Binney]{Carlo Nipoti\thanks{E-mail: nipoti@thphys.ox.ac.uk (CN); binney@thphys.ox.ac.uk (JB)} and James 
Binney\footnotemark[1]
\\
Theoretical Physics, Oxford University,   1 Keble Road, Oxford OX1 3NP, UK\\}
\begin{document}

%\date{Accepted 1988 December 15. Received 1988 December 14; in original form 1988 October 11}

%\date{Submitted 2003 December 2}

\date{Resubmitted 2004 January 8}

\pagerange{\pageref{firstpage}--\pageref{lastpage}} \pubyear{2003}

\maketitle

\label{firstpage}

\begin{abstract}
We determine the critical size $l_{\rm crit}$ of a filament of cold
($T\sim10^4\kelvin$) gas that is in radiative equilibrium with X-ray
emitting gas at temperatures $\tmax \sim 10^6 - 10^8
\kelvin$. Filaments smaller than $l_{\rm crit}$ will be rapidly
evaporated, while longer ones will induce the condensation of the
ambient medium. At fixed pressure $P$, $l_{\rm crit}$ increases as
$\tmax^{11/4}$, while at fixed $\tmax$ it scales as $P^{-1}$. It scales as
$f^{1/2}$, where $f$ is the factor by which the magnetic field
depresses the thermal conductivity below Spitzer's benchmark
value. For plausible values of $f$, $l_{\rm crit}$ is similar to the
lengths of observed filaments.  In a cluster such as Perseus, the
value of $l_{\rm crit}$ increases by over an order of magnitude
between the centre and a radius of $100\kpc$. If the spectrum of seed
filament lengths $l$ is strongly falling with $l$, as is natural, then
these results explain why filaments are only seen within a few
kiloparsecs of the centres of clusters, and are not seen in clusters
that have no cooling flow.  We calculate the differential emission
measure as a function of temperature for the interface between
filaments and ambient gas of various temperatures.  We discuss the
implications of our results for the origin of the galaxy luminosity
function.
\end{abstract}


\section{Introduction}

Intergalactic space within clusters of galaxies contains enormous
quantities of gas that is at the virial temperature. In about three
quarters of all rich clusters, the temperature falls by a factor of 2
to 3 within a few tens of kiloparsecs of the cluster centre -- such
clusters are said to possess a `cooling flow'. Nearby examples are the
Virgo, Hydra and Perseus clusters.

Optical emission-line filaments are observed around NGC 1275 at the
centre of the Perseus cluster (Minkowski 1957; Lynds 1970; Conselice,
Gallagher \& Wyse 2001, and reference therein). Similar structures of
extended $\halpha$ emission are seen near the centres of other
cooling-flow clusters (e.g., Virgo, Hydra, A2597, A2052, A1795; see
Heckman 1981; Hu, Cowie \& Wang 1985; Johnstone, Fabian \& Nulsen
1987; Heckman et al. 1989).  The origin of these filaments and the
mechanism responsible for their ionization have been extensively
studied, but definitive conclusions have not been reached. Two
formation scenarios have been proposed. In the first, the filaments
form as the intracluster medium (ICM) cools (e.g., Fabian \& Nulsen
1977, Cowie, Fabian \& Nulsen 1980). In the other scenario, filaments
are produced through cosmic infall of either cold gas or gas-rich
galaxies (e.g., Soker, Bregman \& Sarazin 1991; Baum 1992; Sparks
1992).  The fact that filaments are observed {\it only\/} in clusters
with cooling flows has often been considered evidence for the first
scenario, but by no means all cooling-flow clusters have detectable
optical emission lines (e.g., A2029; see Fabian 1994).  Some
observational data favour the infall scenario.  For example, Sparks,
Macchetto \& Golombek (1989) showed that filaments contain dust with
normal Galactic extinction properties, which filaments formed by
condensation of the ICM should not (see also Donahue \& Voit 1993).
In addition, the filaments appear to be dynamically disturbed systems,
suggesting that they form through violent processes, such as merging,
rather through the quiescent condensation of the X-ray emitting ICM.

An important related question is, what mechanism is responsible for
heating the filaments? Many possible ionization sources have been
considered: Active Galactic Nuclei (AGN); shocks; massive stars;
radiation from the ICM; magnetic reconnection (e.g., Heckman et
al. 1989). None of these ionization mechanisms, individually, can
satisfactory account for the observed emission-line ratios and
intensities, so possible combinations of these processes have been
explored (e.g., Sabra, Shields \& Filippenko 2000). However, there are
indications that the heating source is physically associated with the
filaments (Conselice et. al 2001 and references therein). Thus, it is
likely that stellar photoionization makes an important contribution to
powering their optical emission-lines, since young stars are
systematically found in the filaments (e.g., McNamara, O'Connell \&
Sarazin 1996).

The existence of cool filaments embedded in hot ambient plasma has
implications for the problem of the plasma's thermal conductivity,
which must be moderated by the intracluster magnetic field. If the
conductivity were high, one would expect the filaments to evaporate
rapidly.  Thus, in principle, their very existence limits the
conductivity in the same way that of cold fronts in the ICM does
(Markevitch et al. 2000, 2003).  Given our very limited understanding
of the connectedness of the intergalactic magnetic field, and the
possibility that thermal conductivity contributes significantly to the
global dynamics of cooling flows (Narayan \& Medvedev 2001, Voigt \&
Fabian 2003 and references therein), observational constraints on
thermal conductivity are valuable.

In this paper we  address two main questions:
\begin{enumerate}
 \item Why are cold filaments observed only in cooling-flow clusters? and
why only in the cluster core?
 \item To what degree does the existence of these filaments constrain
the thermal conductivity of the ICM?
\end{enumerate}
The competition between evaporation and condensation when a cool body
of plasma is immersed in a hotter ambient medium has been studied
extensively in the literature (Graham \& Langer 1973; Cowie \& McKee
1977; McKee \& Cowie 1977; Balbus \& McKee 1982; Giuliani 1984; Draine
\& Giuliani 1984; Balbus 1986; B\"{o}hringer \& Hartquist 1987;
B\"{o}hringer \& Fabian 1989; McKee \& Begelman 1990; Ferrara \&
Shchekinov 1993).  Most effort has been devoted either to spherical
clouds or to planar inhomogeneities, although Cowie \& Songaila (1977)
considered spheroidal clouds.  Since the morphology of observed
filaments suggests that cylindrical symmetry should provide a useful
idealization, we revisit the problem for this case, and discuss the
implications of our results for real systems.

In any geometry, clouds below some critical size are evaporated, while
clouds above this size grow by condensation of the hot medium. We
determine the critical size that separates these two regimes by
finding the steady-state solution in which conductivity exactly
balances radiative losses.  The typical size of observed filaments is
likely to be comparable to the critical size: much smaller filaments
are rapidly evaporated away, while filaments above this critical value
grow slowly, at a rate determined by the cooling time. Significantly
larger filaments are expected to be very rare: in principle, if they
formed they could fundamentally change the nature of the system by
eliminating the cooling hot ambient plasma.

In Section~2 we derive the critical cylindrical
solution. In Section~3 we discuss the application of the model to
cluster filaments. Our conclusions are in Section~4.

\section{Radiatively stabilized cylindrical clouds}

\subsection{Model and equations}

We consider a long cylindrically symmetric cloud in which there is
perfect balance between conductive heat flow towards the symmetry axis
and radiative losses, so the plasma flows neither inwards nor outwards
(see McKee \& Cowie 1977).  The temperature on the axis is assumed to
be $T\sim10^4\kelvin$, while far from the origin $T$ asymptotes to a
value $\sim10^6-10^8 \kelvin$.  With these assumptions, and neglecting
the effects of gravity, the energy equation reads
\begin{equation}\label{eqen} 
-\nabla\cdot{\bf q}=\ne^2\Lambda(T),
\end{equation}
 where ${\bf q}$ is the heat flux, $\ne$ is the electron number density, and
$\Lambda(T)$ is the normalized cooling function. In the unsaturated
thermal conduction regime\footnote{In the Appendix we discuss the
possible effects of saturation.}, the heat flux is given by
 \begin{equation}\label{eqhf}
{\bf q}=-\kappa(T) {\bf \nabla} T, 
\end{equation}
 with $\kappa(T)$ thermal conductivity. For a hydrogen plasma, in the
absence of magnetic fields, Spitzer (1962) derives
$\kappa(T)=\kspitzer T^{5/2}$, with
$\kspitzer\simeq{1.84\times10^{-5}(\ln{\Lambda})}^{-1} \ergs \sm1
\cm^{-1} \kelvin^{-7/2}$, where $\ln\Lambda$ is the Coulomb logarithm,
which is only weakly dependent on $\ne$ and $T$ (in the following we
assume $\ln\Lambda=30$).  Confinement of thermal electrons by the
intracluster magnetic field reduces the actual conductivity to
\begin{equation}\label{eqkappa2} \kappa(T)=f\kspitzer T^{5/2},
\end{equation}
 where the parameter $f \leq 1$ is the heat conduction suppression
factor (e.g., Binney \& Cowie 1981, B\"{o}hringer \& Fabian 1989).

At least in some clusters, observations indicate that emission-line
nebulae are in pressure equilibrium with the surrounding hot plasma
(e.g., Heckman et al. 1989). So we require that the pressure
$\press=\ne T$ is constant throughout the interface, and we use this
relation to eliminate $\ne$ in the right-hand side of
equation~(\ref{eqen}).  Thus, under the assumption of cylindrical
symmetry, eliminating ${\bf q}$ with the help of equations
(\ref{eqhf}) and (\ref{eqkappa2}), the energy equation~(\ref{eqen}) is
reduced to the differential equation
\begin{equation}
\label{eqencyl}
{ f\kspitzer  \over R} {d \over dR} \left( R T^{5/2} {dT \over dR}
\right)=\press^2 { \Lambda(T) \over T^2 },
\end{equation}
 where $R$ is the cylindrical radius.

\subsection{Boundary conditions}

%%%%%%%%%%%%%%FIG 1
\begin{figure}
\centerline{\psfig{file=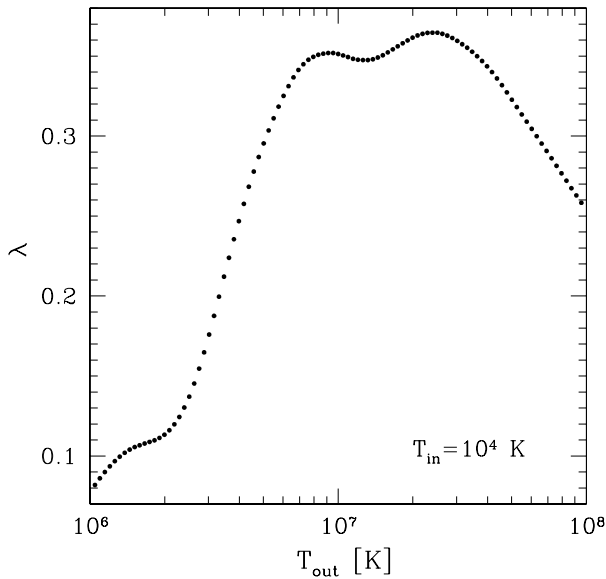,width=\hsize}}
\caption{Eigenvalues $\lambda$ of equation~(\ref{eqxitau}) for a set
of values of the outer temperature $\tmax$.\label{figlam}}
\end{figure}
%%%%%%%%%%%%%%%%%%%%%%%%%%%%%%%%%%%%%%%%%%%%%%%%%%%%%%%%%

Equation (\ref{eqencyl}) applies outside the edge at $\rmin$ of the
$T\sim10^4\kelvin$ cloud: we assume that the cloud contains its own
heat sources, such as photoionizing stars, so on its surface the
temperature gradient and heat flux vanish. We require $T$ to attain
some specified value on $R=\rmax$. Thus we solve equation
(\ref{eqencyl}) subject to three boundary conditions,
$T(\rmin)=\tmin=10^4\kelvin$, $dT/dR(\rmin)=0$, and $T(\rmax)=\tmax$,
the temperature of the ambient hot plasma, which varies with distance
from the cluster centre. For completeness we will explore the wide
range $10^6 \lsim\,\tmax\,\lsim\ 10^8 \kelvin$.

Introducing the dimensionless quantities $\xi=R/\rmax$,
$\tau=T/\tmax$, and $\Lamtmax(\tau)\equiv\Lambda(\tau\tmax)/\Lam23$,
with $\Lam23=10^{-23} \ergs \sm1 \cm^3$, equation~(\ref{eqencyl})
reads \begin{equation}
\label{eqxitau}
{1\over \xi} {d \over d\xi} \left(\xi \tau^{5/2} {d\tau \over d\xi} \right)=
\lambda   { \Lamtmax(\tau) \over  \tau^2},
\end{equation}
 where
 \begin{equation}
\label{eqlambda}
\lambda={\press^2 \rmax^{2} \Lam23 \over f \kappa_0 \tmax^{11/2}}
\end{equation}
 is an unknown dimensionless parameter that is the eigenvalue of
equation (\ref{eqxitau}). Expressed in dimensionless variables, the
boundary conditions are $\tau(\ximin)=\taumin$,
$d\tau/d\xi(\ximin)=0$, and $\tau(1)=1$ with $\ximin \equiv
\rmin/\rmax$ and $\taumin \equiv \tmin/\tmax$.  Due to the presence of
$\Lamtmax$, the eigenfunctions $\tau(\xi)$ and eigenvalues $\lambda$
depend on the outer temperature $\tmax$ in addition to the
dimensionless numbers $\ximin$ and $\taumin$.

The inner and outer radii $\rmin$ and $\rmax$ cannot be exactly identified
with observational quantities.  However, we can assume that $\rmin$
corresponds approximately to the width of the filament, and $\rmax$ to its
length $l$. The latter assumption is consistent with the hypothesis of
cylindrical geometry, which does not hold at $R\gsim l$. Observed filaments
are highly elongated, with ratios length/width $\gsim 100$ (Conselice et al.
2001). Thus, we explore a range of values of $\ximin$ that satisfy $\ximin
\ll 1$.

%%%%%%%%%%%%%%FIG 2
\begin{figure}
\centerline{\psfig{file=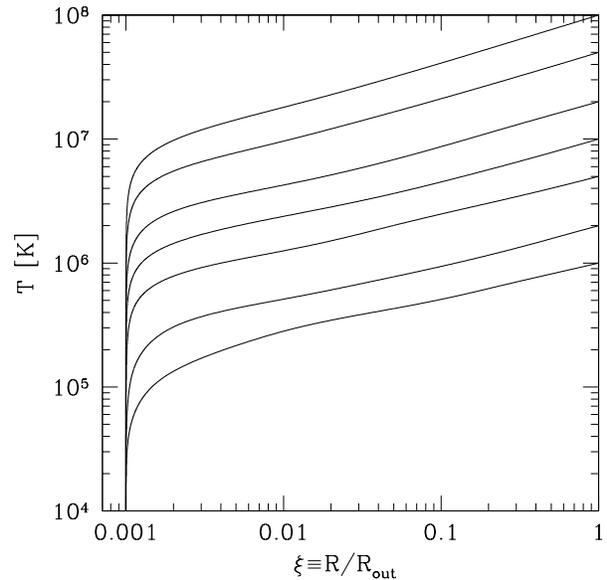,width=\hsize}}
\caption{Temperature profile $T(R)$ corresponding to the eigenfunction
$\tau(\xi)$ of equation~(\ref{eqxitau}), for
$\tmax=0.1,0.2,0.5,1,2,5,10$, in units of $10^7 \kelvin$.\label{figTprof}}
\end{figure}
%%%%%%%%%%%%%%%%%%%%%%%%%%%%%%%%%%%%%%%%%%%%%%%%%%%%%%%%%

\subsection{Numerical solutions}

Equation~(\ref{eqxitau}) is non--linear and would require numerical
integration even if $\Lamtmax(\tau)$ were approximated by a simple
analytical function. Thus, use of the tabulated cooling function of
Sutherland \& Dopita (1993) for metallicity $[{\rm Fe}/{\rm H}]=-0.5$
introduces no additional computational complexity.  We use a shooting
method, with Newton iteration, to calculate the solutions.

Numerical integration of equation~(\ref{eqxitau}) shows that, for any
fixed $\tmax$ in the range considered, the value of $\lambda$ is
virtually independent of $\ximin$, provided that $\ximin$ is
sufficiently small (i.e., $\ximin \lsim 10^{-2}$). Thus, in what
follows we can simplify our treatment by assuming that the eigenvalue
$\lambda$ depends only on $\tmax$.

Although the details of the {\it inner} temperature profile $\tau(\xi)$
depend significantly on $\ximin$, the behaviour of the profile at $\xi
\gsim 10^{-2}$ is virtually independent of $\ximin$ provided $\ximin \lsim
10^{-3}$. Henceforth we set $\ximin \lsim 10^{-3}$ so that {\it both
$\lambda$ and the observationally significant part of $\tau(\xi)$ are
independent on the exact radial range considered}.

Fig.~\ref{figlam} shows $\lambda$ as a function of $\tmax$, while
Fig.~\ref{figTprof} shows a number of the corresponding eigenfunctions
$T(R)$. For values of $\tmax$ in the relevant range, {\it $\lambda$
spans a relatively small range} ($0.08 \lsim \lambda \lsim 0.37$), which
is reduced to $0.25\lsim \lambda \lsim 0.37$ for $\tmax \gsim 4 \times
10^6 \kelvin$.

From Fig.~\ref{figTprof} is apparent that over a significant fraction
of the radial range the temperature profiles are roughly power laws,
with similar indices. This suggests that an approximate analytical
solution could be obtained by exploring the asymptotic behaviour of
the eigenfunctions of equation~(\ref{eqxitau}) for $\xi \to 1$ ($\tau
\to 1$).  When we assume $\tau = \xi^n$, and approximate the cooling function 
near $\tmax$ by $\Lamtmax(\tau)\simeq[\Lambda(\tmax)/\Lam23]
\tau^{\alpha}$, equation~(\ref{eqxitau}) requires $n=4/(11-2\alpha)$,
and $\lambda={7\over2}n^2 / [\Lambda(\tmax)/\Lam23]$.  Since $-1/2 \lsim
\alpha \lsim 1/2$ for the relevant temperatures, the outer slope of
the eigenfunction is expected to lie in the small range $1/3 \lsim n
\lsim 2/5$ (cf. Fig.~\ref{figTprof}). The run of the eigenvalue as a
function of $\tmax$ (Fig.~\ref{figlam}) reflects the proportionality
to $1/\Lambda(\tmax)$, modulated by $n^2$, which accounts for the
dependence on the local slope of the cooling function $\alpha$.

\section{Applications}

\subsection{Critical length for filaments}

We now discuss the physical implications of the results presented
in the previous section. Equation~(\ref{eqlambda}) relates the numerical value of $\lambda$
to the dimensional quantities associated with a filament and its
surroundings. If, as explained above, we identify $\rmax$ with the length of a
filament, then this equation yields for a given environment the critical length $\lcrit$ of
filaments, such that smaller filaments evaporate while larger ones grow.
Numerically we have
 \begin{eqnarray}\label{eqrmaxtmax}
\lcrit&\equiv&\rmax\nonumber\\
&\simeq& 1.4f^{1/2} \lambda^{1/2}
\left({\tmax \over 10^7 \kelvin }\right)^{11/4}
\left({\press \over 10^6 \kelvin \cmm3}\right)^{-1}{\kpc}.
\end{eqnarray}
 This equation indicates that $\lcrit$ has a strong dependence on the
outer temperature ($\lcrit \propto \tmax^{11/4}$, neglecting the weak
dependence of $\lambda$ on $\tmax$; see Fig.~\ref{figlam}), and also
is inversely proportional to the pressure $\press$.  Even though the
dependence of $\lcrit$ on the thermal conduction suppression factor
$f$ is not very strong ($\lcrit \propto f^{1/2}$), $f$ can
significantly affect $\lcrit$ because the value of $f$ is very
uncertain: $f\sim1$ seems to be excluded, but different authors gives
values differing by two orders of magnitude, from a few $10^{-1}$ to
$\sim 10^{-3}$ -- see Narayan \& Medvedev (2001) for a discussion. We
consider values in the range $10^{-3} \lsim f \lsim 1$ and discuss the
possibility of constraining $f$ from observations of filaments in
Section~3.4.

%%%%%%%%%%%%%%FIG 3
\begin{figure}
\centerline{\psfig{file=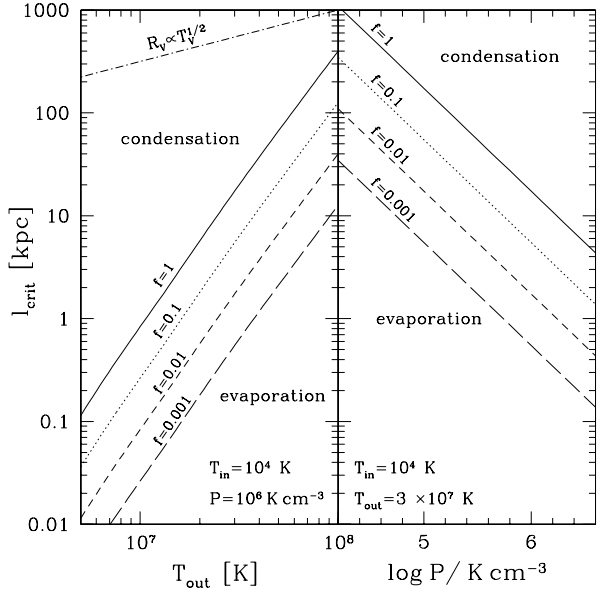,width=\hsize}}
\caption{{\it Left}: Critical length of a radiatively stabilized
filament as a function of the temperature of the ambient medium, at
fixed pressure $P=10^6 \kelvin \cm^{-3}$, for a few values of the
suppression factor $f$. The dot-dashed line represents the
self-similar relation between virial temperature and virial
radius. {\it Right}: Critical length as a function of pressure, at
fixed temperature $T=3 \times 10^7 \kelvin$, for the same set of
values of $f$ as in left panel.\label{figrmaxtmax}}
\end{figure}
%%%%%%%%%%%%%%%%%%%%%%%%%%%%%%%%%%%%%%%%%%%%%%%%%%%%%%%%%

Consider the implications of the dependence of $\lcrit$ on $\tmax$ at
fixed pressure.  The left panel of Fig.~\ref{figrmaxtmax} shows
$\lcrit(\tmax)$ from equation~(\ref{eqrmaxtmax}), for pressure
$\press=10^6 \kelvin \cm^{-3}$, which is a typical value in the core
of a cooling-flow galaxy cluster [e.g., for the Perseus cluster,
Conselice et al.\ (2001), Schmidt, Fabian \& Sanders\ (2002); see also
Heckman et al.\ (1989)].  A few values of the suppression factor
($f=1,0.1,0.01,0.001$) are considered. For given $f$, the relation
$\lcrit(\tmax)$ divides the the diagram in two regions: filaments
longer than $\lcrit$ are condensing, while smaller ones are
evaporating. It is apparent from the steep slope of the curves in the
diagram that the process of evaporation or condensation of filaments
is very sensitive to the temperature of the ambient medium. For a
plausible value of the suppression factor, $f\sim 0.01$, at fixed
pressure $\sim10^6 \kelvin \cm^{-3}$, any filament longer than $\sim
0.1 \kpc$ is condensing in a cluster with ambient temperature
$\tmax=10^7 \kelvin$, while filaments as long as $\sim 10 \kpc$ are
evaporating if $\tmax=6 \times 10^7 \kelvin$.

Although $\lcrit$ depends on pressure more weakly than on temperature,
pressure also plays an important role in determining the fate of
filaments. This is apparent in the right diagram of
Fig.~\ref{figrmaxtmax}, where we fix $\tmax=3 \times 10^7$ and plot
$\lcrit$ as a function of $P$, for the set of values of $f$ considered
above. The diagram shows that, at fixed temperature, in low-pressure
clusters virtually all filaments are evaporating (e.g., $\lcrit\sim
100 \kpc$ for $\press \sim {\rm few} \times 10^4 \kelvin \cm^{-3}$ and
$f=0.01$), but condensation is effective in the highest-pressure
ambient gas ($\lcrit \sim 1 \kpc$ for $P \gsim 10^6 \kelvin \cm^{-3}$
and $f=0.01$).

\subsection{Time-scales}

What are the characteristic time-scales for evaporation and
condensation of filaments? A detailed answer to this question would
require the solution of the full hydrodynamic equations for
non-vanishing mass flow. However, for the purpose of our investigation
we are interested in an order of magnitude estimate, which can be
obtained by considering that the time variation of temperature is
described by a differential equation of the form
\begin{equation}\label{eqdTdt} 
{ dT \over dt} = { T \over \tkappa} - { T \over \tcool},
\end{equation}
where $\tkappa$ and $\tcool$ are the thermal conduction and cooling
time-scales, respectively.  The relation between these time-scales is
given by 
\begin{equation}\label{eqtktc} 
{ \tkappa \over \tcool}=
{\ne^2\Lambda(T) \over |\nabla\cdot{\bf q}|}
\sim {\ne^2\Lambda(T) l^2  \over \kappa(T)T},
\end{equation}
because $|\nabla\cdot{\bf q}|\sim \kappa(T)T/l^2$, where $l$ is the
temperature variation length scale (Begelman \& McKee 1990). In the
considered case, $l$ is the radial extension of the interface, which is
of the order of the filament length (see Section~2). Considering that
the equilibrium solution ($\l=\lcrit$) corresponds to
$\tkappa\sim\tcool$, from equation~(\ref{eqtktc}) we get
\begin{equation}\label{eqtk} 
{ \tkappa \sim {\left({ l \over \lcrit}\right)^2  \tcool}},
\end{equation}
and equation~(\ref{eqdTdt}) can be approximated as
\begin{equation}\label{eqdTdt2} 
{ dT \over dt} = { T \over \tcool}\left[{\left({ \lcrit \over \l}\right)^2-1}\right].
\end{equation}
 Consequently, filaments shorter than $\sim \frac12l_{\rm crit}$
evaporate in a fraction $(l/l_{\rm crit})^2$ of the cooling time,
which for example is $\lsim300\Myr$ within $\sim4\kpc$ of the centre
of the Hydra cluster, (David et al.\ 2000). Filaments smaller than
$l_{\rm crit}$ by a factor $\gsim8$ have evaporation times
$\lsim5\Myr$ that are smaller than their likely dynamical times,
$l/v$, where $v\sim100\kms$ is the typical gas velocity estimated from
line widths (Sabra et al. 2000 and references therein). On the other
hand, equation~(\ref{eqdTdt2}) in the limit $l\gg \lcrit$ indicates
that, as expected, very long filaments condense on time-scales of the
order of the cooling time of the ambient medium.

These considerations strongly support the hypothesis that {\it only
filaments with lengths comparable to the critical length or longer are
likely to survive for long enough to be observed}.

\subsection{The fate of filaments in cooling-flow and non--cooling-flow clusters}

%%%%%%%%%%%%%%FIG 4
\begin{figure}
\centerline{\psfig{file=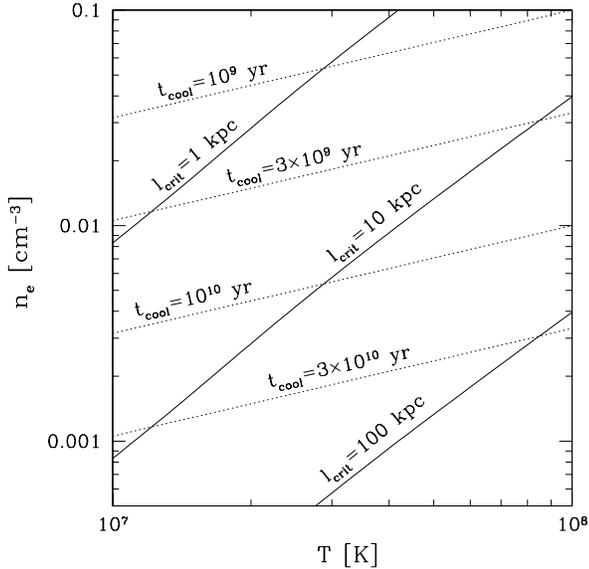,width=\hsize}}
\caption{Dotted curves: density--temperature relation for plasma with
fixed cooling time ($\tcool=0.1,0.3,1,3$ in units of $10^{10}$ yr).
Solid curves: density as function of temperature for $\lcrit=1,10,100
\kpc$ and $f=0.01$ (equation~\ref{eqrmaxtmax}).\label{fignT}}
\end{figure}
%%%%%%%%%%%%%%%%%%%%%%%%%%%%%%%%%%%%%%%%%%%%%%%%%%%%%%%%%

The results presented above have interesting implications, when
compared with the properties of observed galaxy clusters.

\begin{enumerate}
 \item A first simplistic implication is that, due to the strong
dependence of $\lcrit$ on temperature, in massive (high-temperature)
clusters most cold gas structures are evaporating, while even
relatively small clouds can survive (and condense) in small
(low-temperature) clusters. On the other hand one needs to bear in
mind that clusters with higher temperatures have larger virial radii
$R_{\rm V}$, and it is natural to expect the typical size of clouds to
increase proportional to $R_{\rm V}$.  The left panel of
Fig.~\ref{figrmaxtmax} shows the relation between $R_{\rm V}$ and the
virial temperature ($R_{\rm V} \propto T_{\rm V}^{1/2}$) predicted for
galaxy clusters under the assumption of self-similarity (the slope of
this scaling relation is consistent with that derived from X-ray
observations of galaxy clusters; see, e.g., Ettori, De Grandi \&
Molendi 2002, and references therein). Thus, as is apparent from the
diagram, the relation $\lcrit(\tmax)$ is much steeper than the
relation between cluster size and temperature.  Hence, in the
highest-temperature clusters any cold filament is evaporating, while
in the lowest-temperature clusters, any resolvable filament is
condensing.

\item The scenario presented in the previous point would hold under
the assumption that the physical properties (gas density and pressure)
of clusters of given temperature would not differ significantly from
cluster to cluster. In fact, we know this is not the case: for fixed
gas temperature, non--cooling-flow clusters (i.e., ones with long
cooling times) have, by definition, lower central density (and
pressure) than cooling-flow clusters.  Thus, as shown by the right
panel of Fig.~\ref{figrmaxtmax}, since $\lcrit \propto 1/P$, a
filament long enough to be condensing in the core of a cooling-flow
cluster with central temperature $T$ might evaporate if it were at the
centre of a non--cooling-flow cluster with the same temperature (but
lower density).  This argument suggests that filaments of cold gas are
not observed in non--cooling-flow clusters because in these clusters
they are evaporated very efficiently, even at the cluster centre. On
the other hand, whatever their formation mechanism, filaments are more
likely to survive and grow in the denser, higher pressure cores of
cooling-flow clusters.

\item In a given cooling-flow cluster, $T$ increases as one moves away
from the centre, while $P$ falls. Consequently, equation
(\ref{eqrmaxtmax}) predicts that $\lcrit$ increases rapidly with
distance from the cluster centre. This result explains why, even in
the infall or merging scenario for the formation of the filaments,
these are observed only in the cooling cores of clusters, and not in
the outer regions.  For example, in the Perseus cluster, the gas
temperature at a distance $\sim100 \kpc$ from the center is roughly
twice the central temperature. Also, in the same radial range the
pressure decreases by a factor of $\sim 2.5$ (e.g., Schimdt et
al. 2002; Churazov et al. 2003). Thus, equation~(\ref{eqrmaxtmax})
predicts that $\lcrit$ grows by over a factor of 10 between the centre
and $100\kpc$. Interestingly, even in the core, observations indicate
that the mean $\halpha$ emission is a strongly decreasing function of
the cluster radius (e.g., Conselice et al. 2001). This result may
arise because as one moves out, $\halpha$ emission from smaller
filaments contributes less and less to the overall signal.

\end{enumerate}

We can summarize the points above by considering the relation between
$\lcrit$ and the cooling time of ambient gas.  For simplicity, we use
the approximate isobaric cooling time $\tcool \simeq 10^{4} \ne^{-1}
T^{1/2}\,{\rm yr}$ (where $\ne$ is in $\cm^{-3}$ and $T$ in $\kelvin$;
e.g., Sarazin 1986), considering the temperature range $10^7-10^8
\kelvin$.  The dotted curves in Fig.~\ref{fignT} represent the loci of
constant cooling time in density-temperature space, while the solid
curves show the loci of constant $\lcrit$, for suppression factor
$f=0.01$. We see that if the cooling time is long (as in the cores of
non cooling-flow clusters, and in the outer regions of any cluster),
$\lcrit$ is large, so evaporation is very likely. On the other and,
small $\lcrit$ corresponds to a short cooling time. Hence {\it it is
precisely the physical conditions found in the cores of cooling-flow
clusters that favour condensation}.

%%%%%%%%%%%%%%FIG 5
\begin{figure}
\centerline{\psfig{file=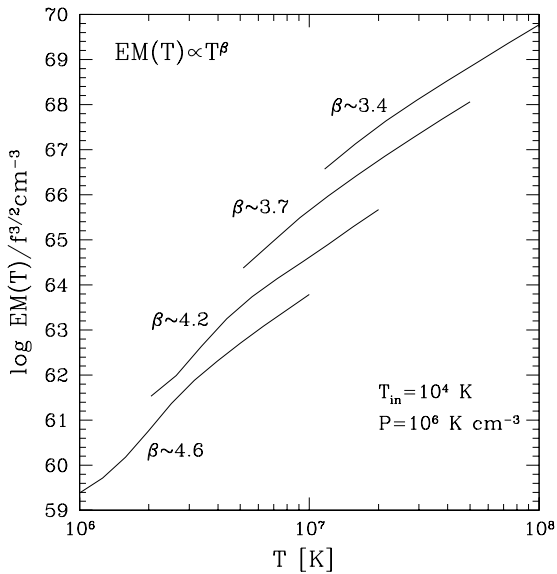,width=\hsize}}
\caption{Differential emission measure $EM(T) \equiv \ne^2 dV /d \log T$
as a function of $T$, corresponding to the critical stationary
solution for outer temperatures $\tmax=1,2,5,10$, in units of $10^7
\kelvin$. $f$ measures the reduction of heat conduction below
Spitzer's value. $\beta$ is the index of the best--fitting power law.
\label{figEM}}
\end{figure}
%%%%%%%%%%%%%%%%%%%%%%%%%%%%%%%%%%%%%%%%%%%%%%%%%%%%%%%%%

\subsection{Suppression of thermal conductivity in the intracluster medium}

As outlined in the Introduction, the existence of cold filaments could
be used to constrain the value of the thermal conductivity in the ICM.
Indeed, we could derive the value of the suppression factor $f$ from
equation~(\ref{eqrmaxtmax}) if we knew all the other quantities
involved.  Unfortunately, while $\tmax$ and $P$ are well constrained
by observations, it is quite difficult to put strong observational
constraints on $\lcrit$. A detailed morphological study of the
filaments, available only for the Perseus cluster (Conselice et
al. 2001), reveals complex structures on different scales. However, we
note that even a quite conservative estimate of $\lcrit$ can lead to
interesting constraints on $f$. For instance, the data suggest that in
the Perseus cluster filaments $\sim 10 \kpc$ long are present at least
up to radii $\sim 40 - 50 \kpc$.  This means $\lcrit \lsim 10 \kpc$,
where the ICM temperature and pressure are $\sim 4.5 \times 10^7
\kelvin$ and $\sim 10^6 \kelvin \cm^{-3}$, respectively (Schmidt et
al.  2002).  According to equation~(\ref{eqrmaxtmax}), for these
physical properties of the ambient medium, this condition is
translated into the {\it upper limit} $f \lsim 0.04$ for the thermal
conduction suppression factor. This result is compatible with that
found by Markevitch et al. (2003) on the basis of observations of cold
fronts in A754.  We note that a substantially higher efficiency would
be required in order for the thermal conductivity to significantly
affect cooling flows in the centres of clusters (e.g., Narayan \&
Medvedev 2001).

In the discussion above we considered the thermal conduction
suppression factor as a global parameter characterizing the ICM in
clusters. In fact, the reduction of thermal conductivity could vary
from cluster to cluster, or within a given cluster, depending on the
detailed properties of the magnetic field. However, there are
indications that the effect of the reconnection of the magnetic field
lines between the filaments and the ICM would be to enhance the
efficiency of heat conduction in the centres of cooling-flow clusters
(Soker, Blanton \& Sarazin 2004). Thus, at least according to this
scenario, it seems reasonable to extend the derived upper limit on $f$
to the ICM in general.

\subsection{X-ray emission from conduction fronts of filaments}

Here we explore the possibility that the conduction fronts described
above can be detected in soft X-rays. This is particularly interesting
since Fabian et al. (2003) find soft X-ray emission associated with
the brightest $\halpha$ emission filaments around NGC1275, suggesting
that it is produced in the conductive evaporation or condensation of
the filaments.  We quantify the emission properties of the conduction
fronts for the considered cylindrically symmetric clouds, computing
the differential emission measure
\begin{equation}\label{eqemone} 
EM(T) \equiv \ne^2 {dV \over
d{\log}T},
\end{equation}
 where $\ne$ is the electron number density and $dV$ is the volume
element.  Under the hypothesis of constant pressure $P=\ne T$, the
differential emission measure can be related directly to the
temperature profile $T(R)$: using $dV=2\pi l R dR$, with $l$ length of
the filament, we get
\begin{equation}\label{eqemtwo}
EM(T) = 2 \pi l P^2 {R \over T^2} {dR \over d{\log}T}.
\end{equation}
As discussed in Section~2, the temperature profile, for given $\tmax$,
can be considered independent of $\ximin$ at large enough radii ($\xi
\gsim 10^{-2}$); correspondingly we find that the differential
emission measure at $T\gsim 0.1 \tmax$ is not sensitive to the value
of $\ximin$.  Fig.~\ref{figEM} plots the emission measure as a
function of $T$ for a subset of values of the outer temperature
($\tmax=1,2,5,10$ in units of $10^7 \kelvin$).  For the normalization
of $EM(T)$ we assume $P=10^6 \kelvin \cm^{-3}$, and $l=\lcrit=\rmax$,
in accordance with the discussion in Section~2. Due to the dependence
of $\lcrit$ on the thermal conduction suppression factor $f$ (equation
\ref{eqrmaxtmax}), the normalization of $EM(T)$ is proportional to
$f^{3/2}$. Thus it is difficult to make predictions on the absolute
value of the emission measure, also because of the strong dependence
on pressure.  On the other hand, the shape of $EM(T)$ is not affected
by these parameters, being determined by the temperature profile,
which depends only on the temperature of the ambient medium. As
apparent from Fig.~\ref{figEM}, in the considered temperature range
the differential emission measure can be roughly approximated with a
power-law $EM(T)\propto T^{\beta}$. We find that $EM(T)$ is steeper
for lower $\tmax$, with best--fitting slope $\beta$ ranging from
$\beta\simeq3.4$ for $\tmax=10^8 \kelvin$ to $\beta\simeq4.6$ for
$\tmax=10^7 \kelvin$. Accurate measurements of the soft X-ray spectra
around filaments could be interestingly compared with the slopes of
these emission measure profiles.

\section{Conclusions}

Embedded within cooling flows, one frequently detects filaments of cold gas.
Thermal conduction will cause heat to flow into these filaments from the hot
ambient gas, with the result that filaments below a critical size will be
evaporated. On the other hand radiative cooling is very efficient in cold,
dense gas, so larger filaments can survive, and indeed nucleate the cooling
of the ambient medium. We have determined the critical filamentary size that
divides the two regimes.

The absolute value of this length scale, $l_{\rm crit}$, is uncertain
because we do not know how effectively the magnetic field suppresses
thermal conduction.  By contrast, the way in which $l_{\rm crit}$
scales with the temperature and pressure of the ambient medium is well
determined and leads to suggestive results.

The critical length scale rises sharply with ambient temperature:
$l_{\rm crit}\propto T^{11/4}$, and decreases as $P^{-1}$ with ambient
pressure.  Consequently, in the Perseus cluster $l_{\rm crit}$
increases by over an order of magnitude between the cluster centre and
a radius of $100\kpc$, where it is probably larger than
$10\kpc$. Within a few kiloparsecs of the cluster centre, filaments
only a kiloparsec long may either grow, or at least take a long
time to evaporate because their lengths are just below the critical
value, while such small filaments are rapidly evaporated in the main
body of the cluster.

Where do filaments come from in the first place? They cannot emerge
spontaneously from the ambient medium because this is thermally
stable,\footnote{Balbus (1991) shows that thermal instability is
possible in the presence of even a weak magnetic field. However, a
weak field can only control low-amplitude perturbations, and buoyant
stabilization will reassert itself once the perturbation amplitude is
greater than $P_{\rm mag}/P_{\rm gas} \sim 10^{-3}$ in clusters. It is
also the case that for technical reasons Balbus had to insist on a
highly artificial background field for which field-line closure is
problematic.} so they must be injected.  Simulations of structure
formation predict that significant amounts of cold gas fall into
clusters (Katz et al. 2003), and the observed filaments could
represent just the fraction of this infall that reaches the centre.
There is also some indication that extremely close to the cluster
centre a small amount of gas cools to filamentary temperatures, and
some of this gas could be ejected by the AGN to the radii where
filaments are observed.

Whether infall or AGN are responsible for injecting filamentary gas,
there is no reason to expect the lengths of the injected filaments to
have a characteristic value.  It is natural to think in terms of a
spectrum of filamentary lengths $l$, in which the number of filaments
decreases with increasing filament length, probably as a power law.
Observations of filaments would then be dominated by filaments with
$l\sim l_{\rm crit}$ because shorter filaments would be rapidly
evaporated, and larger ones would be rare. The local number density of
filaments would be a decreasing function of $l_{\rm crit}$, explaining
why observed filaments are concentrated near the centres of
cooling-flow clusters.

Since filament growth depends on the efficiency of radiative cooling,
$l_{\rm crit}$ tends to be small when the cooling time of the ambient
medium is short. Hence clusters with short central cooling times, have
smaller values of $l_{\rm crit}$ and more filaments than clusters with
similar temperatures but longer cooling times. Consequently, it is
natural that filaments are only detected in cooling-flow clusters.

For the case of a filament of critical length, we have calculated the
temperature profile throughout the interface between gas at
$\sim10^4\kelvin$ that emits optical emission lines, and the ambient
X-ray emitting plasma. This profile yields the run of differential
emission measure with temperature, which can be tested
observationally.

As cosmic clustering proceeds, the temperature of gas that is trapped
in gravitational potential wells rises as longer and longer
length-scale perturbations collapse. This temperature rise causes
$l_{\rm crit}$ to rise much faster than the virial radius of the
potential well.  Hence, while at early times filaments that are much
smaller than the system can grow, growth is later restricted to
filaments that are comparable in size to the entire
system. Consequently, the ability of infalling cold gas to nucleate
star formation declines as clustering proceeds.

Semi-analytic models of galaxy formation that can account for the
observed slope of the galaxy luminosity function at the faint end need
efficient feedback from star formation, and over-produce luminous
galaxies (Benson et al. 2003).  Binney (2003) has argued that galaxy
formation is dominated by infall of cold gas, and that it is the
efficient evaporation of filaments in massive systems that suppresses
the formation of massive galaxies. Our results make a step towards
quantifying this picture.

\section*{Acknowledgments}

We thank Luca~Ciotti for helpful comments on the manuscript, and the
referee, Noam~Soker, for interesting suggestions that helped improve
the paper. C.N. was partially supported by a grant from Universit\`a
di Bologna.

\appendix

\section[]{Thermal conduction saturation parameter}

We have hitherto assumed that the thermal conduction is not saturated,
so the heat flux is given by the classical equation~(\ref{eqhf}), with
the thermal conductivity given by equation~(\ref{eqkappa2}). The
latter formula is derived under the assumption that $l_e/l_T \ll 1$,
where $l_e$ is the electron mean free path and $l_T=T / |\nabla T|$ is
the temperature scale height. When this condition is violated,
equation~(\ref{eqhf}) overestimates the heat flux, which is limited by
the thermal speed of electrons, and the modulus of ${\bf q}$ is given
by the saturated heat flux (Cowie \& McKee 1977)
\begin{equation}
\label{eqqsat}
\qsat=0.4 \left({ 2 \kb T \over \pi \me} \right)^{1/2} \ne \kb T,
\end{equation}
 where $\kb$ is the Boltzmann constant, and $\me$ is the electron
mass.  Cowie \& McKee introduced a discriminant of whether the heat
flux in a plasma is saturated: this is the {\it saturation parameter}
$\sigma \equiv \qcl/\qsat$, where $\qcl=\kappa(T) |{\bf \nabla} T|$ is
the modulus of the classical heat flux (equation~\ref{eqhf}).  The
critical value that separates the two regimes is $\sigma \sim 1$: for
$\sigma \ll 1$ the heat flux is unsaturated, while for $\sigma \gg 1$,
equation~(\ref{eqqsat}) applies.

For given $\tmax$ and corresponding critical temperature profile
$T(R)$, the value of the saturation parameter as a function radius in
the interface can be obtained combining equations~(\ref{eqhf}),
(\ref{eqkappa2}), and (\ref{eqqsat}), and using the relations $\ne=P/T$
and $\rmax=\lcrit$ (equation~\ref{eqrmaxtmax}). Numerically we get
\begin{equation}\label{eqsigma} 
\sigma (R) \simeq 8 \times 10^{-3} f^{1/2} \lambda^{-1/2} \left({\tmax
\over 10^7 \kelvin}\right)^{1/4}\tau^2 {d\tau \over d\xi},
\end{equation}
where $\xi\equiv R/\rmax$, $\tau=\tau(\xi)$ is the normalized
temperature profile, and $\lambda=\lambda(\tmax)$ the corresponding
eigenvalue (see Section~2). We note that $\sigma(R)$ is independent of
the pressure and scales as the square root of the suppression factor
$f$. Applying equation~(\ref{eqsigma}) to conduction fronts with
ambient gas temperature in the considered range $10^6
\lsim\,\tmax\,\lsim\ 10^8 \kelvin$, we find in all cases $\sigma(R)\lsim
10^{-2}f^{1/2}$. Thus, the condition for unsaturated heat conduction
($\sigma \ll 1$) is satisfied throughout the interface, even for the
largest values of the suppression factor $f \sim 1$.

\bsp

\label{lastpage}

\end{document}